\documentclass[aps,prb,showpacs,twocolumn,groupedaddress]{revtex4-1}

\usepackage{amsmath}
\usepackage{amssymb}
\usepackage{graphicx}
\usepackage[english]{babel}
\usepackage{blindtext}
\usepackage[utf8]{inputenc}
\usepackage{braket}
\usepackage{bbold}

\usepackage{subfigure}

\usepackage{MnSymbol}

\newcommand{\CaO}{CaO}
\newcommand{\AlTwoOThree}{Al$_{\mathrm{2}}$O$_{\mathrm{3}}$}

\begin{document}


\title{Surface mode hybridization in the optical response of core-shell particles}


\author{E. Thiessen, R. L. Heinisch, F. X. Bronold, and H. Fehske}
\affiliation{Institut für Physik, Ernst-Moritz-Arndt-Universität Greifswald, 17489 Greifswald, Germany}


\date{\today}

\begin{abstract}
We present an exact rewriting of the Mie coefficients describing the scattering of light 
by a spherical core-shell particle which enables their interpretation in terms of an 
hybridization of the two surface modes arising, respectively, at the core-shell and the 
shell-medium interface. For this particular case we thus obtain from the Mie 
theory--analytically for all multipole orders and hence for arbitrarily sized particles--the 
hybridization scenario, which so far has been employed primarily for small particles 
in the electrostatic approximation. To demonstrate the strength of the rewriting 
approach we also extract the hybridization scenario for a stratified sphere directly
from the expansion coefficients for the electromagnetic fields.
\end{abstract} 

\pacs{42.25.Bs, 42.25.Fx}

\maketitle

\section{Introduction}
Ever since the pioneering work by Mie~\cite{Mie08} and Debye~\cite{Debye09}, the classical 
optical response of objects plays an important role in applied science~\cite{Kerker69,BH83}. 
The applications range from the spectroscopy of grains embedded in gaseous interstellar
environments~\cite{HT74} to plasmonic devices on the nanoscale~\cite{MA05,FZ14}. In particular 
the latter is a growing field of research driven by the progress in materials synthesis and 
processing~\cite{KMH14} which provides plasmonic nanostructures of continuously increasing 
complexity~\cite{HLL12,PS12}: nanorice~\cite{WB06}, nanorings~\cite{AH03}, or 
nanoshells~\cite{OA98,BG10} to name only a few. The geometry of these structures provides 
efficient means to tailor their optical response, opening thereby new fields of application, 
for instance, in biomedicine~\cite{LL04} or photovoltaics~\cite{CH06}, which in turn stimulates 
the design of even more intricate structures.

Different methods have been developed over the years to analyze light scattering by composite 
objects. While the approaches~\cite{AK51,W62,R75,U80,B85,SQ94,X05,PP09} extending the early 
analytical works~\cite{Mie08,Debye09} are usually restricted to spherical particles, numerical 
approaches are now available which do not suffer this limitation~\cite{W98,ZVK12,Khlebtsov13}. 
Depending on the discretization, the methods are either surface- or volume-based. In particular 
the latter are very powerful since they can handle arbitrarily shaped inhomogeneous objects. But 
like the generalizations of Mie's original approach numerical methods provide no physical 
picture of the interaction of light with composite objects.

An appealing physical picture was first given by Prodan and coworkers~\cite{Prodan03,PN04} for 
metallic nanoshells and later by Preston and Signorell~\cite{Preston11} for dielectric core-shell 
particles. As pointed out by these authors the optical response of composite entities such 
as nanoshells can be understood within the surface mode hybridization scenario. The essence 
of the scenario is that the surface modes arising at the interfaces of these objects interact 
with each other and result in optical resonances not to be found in homogeneous objects. Modifying 
the object's geometry changes the interaction and hence the optical response. Prodan and coworkers 
showed that the interaction can be interpreted as a kind of hybridization. In 
analogy to electronic states of condensed matter, optical resonances of composite objects can thus 
be labelled bonding and antibonding depending on their symmetry. Both Prodan and coworkers~\cite{Prodan03} 
and Preston and Signorell~\cite{Preston11} worked out the hybridization scenario within the electrostatic 
approximation using, respectively, a Lagrangian and an eigenvalue method for the description of the 
charges induced at the interfaces of the object. The two approaches are very flexible, can be applied 
to arbitrary geometry, and require not necessarily bulk dielectric functions. They can be combined with 
microscopic models for the optical response of the atomic constituents of the objects and are thus well 
suited for applications in nanoplasmonics~\cite{HLL12,PS12}.  

The interpretation of light scattering by composite objects in terms of the hybridization scenario 
is often referred to as a new conceptual approach~\cite{RH04} or as a fundamentally different way 
of thinking about plasmonic effects~\cite{JM11}. As the Mie theory is the exact analytical 
description of light scattering by spheres and thus naturally includes all observed effects, the 
question arises whether the hybridization scenario can be also found directly in the Mie formulae 
and thus whether the new thinking can be united with the old formalism.

In this paper we show that this is indeed the case. For the particular case of a spherical 
core-shell particle the hybridization scenario can be straightforwardly derived from the Mie theory 
by recasting the Mie coefficients~\cite{Kerker69,BH83,AK51,TBH15} in a form resembling the diagonal 
elements of a matrix resolvent describing two hybridized energy levels. The derivation, 
valid for all multipole orders and thus applicable to arbitrarily sized particles, starts with a 
splitting of the electromagnetic fields inside the shell into two parts~\cite{X05,PP09}, corresponding 
to the penetrating fields of an homogeneous particle made out of shell material and the scattering 
fields of a cavity filled with core material and sitting inside an homogeneous domain of shell 
material. In a second step the expansion coefficients obtained by enforcing the boundary conditions 
at the core-shell and shell-medium interface are then rewritten into the desired form.  
The re-organization of the fields inside the shell and the re-ordering of the expansion coefficients 
it leads to yields no physical concepts beyond the hybridization scenario~\cite{Prodan03,PN04,Preston11}. 
In fact it was inspired by it. But it is comforting to see the scenario emerge directly from the 
formulae of the Mie theory, which are notoriously difficult to interpret. We thereby also generalize 
the work of Ruppin~\cite{R75} and Uberoi~\cite{U80} who showed, again only in the electrostatic 
approximation, that the optical response of a core-shell particle arises from two coupled subsystems. 
The hybridization scenario can be also found in the Mie formulae for a stratified sphere. Essential 
is again the splitting of the fields into penetrating and scattered parts and the re-organization of 
the expansion coefficients guided by the structure of the diagonal elements of a matrix resolvent 
describing hybridized energy levels.

In the next section we rewrite the Mie coefficients for a spherical core-shell particle and demonstrate 
how the building blocks of the hybridization scenario, the medium-embedded homogeneous shell particle 
and the shell-embedded core cavity, can be extracted from the new expressions. For illustration we 
present in Sect.~\ref{Results} numerical results for a dielectric core-shell particle, the type of 
particle we suggested to employ in a gas discharge as an electric probe with optical read-out~\cite{THB14}. 
In particular we show, up to the hexapole, data for the splitting between the bonding and antibonding 
resonances as well as the electric fields inside the particle. The stratified spherical particle is 
discussed separately in an Appendix. Concluding remarks are given in Sec.~\ref{Conclusions}.


\section{Theory} \label{Theory}
\begin{figure}[t]
\includegraphics[width=0.6\linewidth]{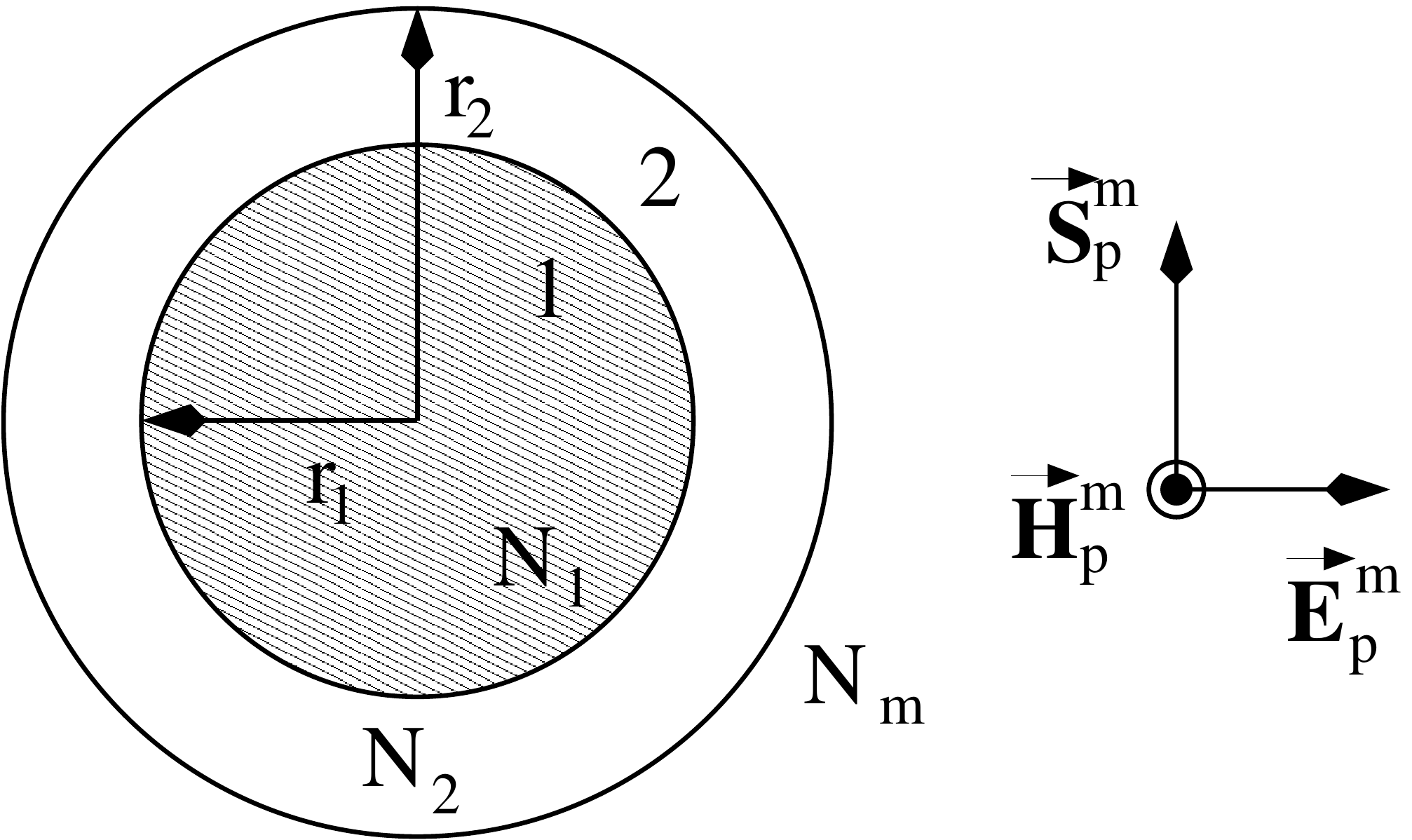}%
\caption{\label{fig0} Geometry of light scattering by a core-shell particle with total radius $r_2$
and core radius $r_1$ embedded in a medium. Refractive indices $N_i$ characterize the core ($i=1$),
the shell ($i=2$), and the medium $i=m$. The incident electromagnetic plane wave is described by a 
Poynting vector $\vec{S}_{\rm in}$ propagating in $z$ direction, an electric field $\vec{E}_{\rm p}^m$ 
polarized in $x$ direction, and a magnetic field $\vec{H}_{\rm p}^m$ along the $y$ direction.}
\end{figure}

The Mie theory~\cite{Mie08,Debye09} of light scattering by a spherical core-shell particle has been 
worked out by Aden and Kerker~\cite{AK51} long time ago. It can be found in many
textbooks~\cite{Kerker69,BH83}. The geometry of the electromagnetic scattering problem is shown in 
Fig.~\ref{fig0}. An electromagnetic wave with wave number $\lambda^{-1}$ propagating in $z$ direction 
and an electric field polarized in $x$ direction hits a particle centered in the origin of the 
coordinate system. The particle with total radius $r_2$ and refractive index 
$N_2(\lambda^{-1})=\sqrt{\varepsilon_2(\lambda^{-1})}$ contains a core with radius $r_1=fr_2$ and  
refractive index $N_1(\lambda^{-1})=\sqrt{\varepsilon_1(\lambda^{-1})}$, where 
$\varepsilon_{1,2}(\lambda^{-1})=\varepsilon_{1,2}^\prime(\lambda^{-1})+i\varepsilon_{1,2}^{\prime\prime}(\lambda^{-1})$
are the complex dielectric functions for the two regions and $0\le f\le 1$ is the filling factor. In the  
formulae below we use $\kappa=2\pi\lambda^{-1}$ and $\omega=2\pi c \lambda^{-1}$ instead of $\lambda^{-1}$
and $\nu$ with $c$ the speed of light, the abbreviations $k_i=\kappa N_i$ with $i=1,2,m$, where $N_m$ is
the refractive index of the surrounding medium, and the size parameters $x_2=2\pi\lambda^{-1} r_2$ and 
$x_1=2\pi\lambda^{-1} r_1$.

In order to obtain the classical optical response of the particle we expand, as in the standard 
procedure, the electromagnetic fields outside and inside the particle in the vector spherical 
harmonics $\vec{M}, \vec{N}$ and determine the expansion coefficients from the boundary conditions 
at the two interfaces~\cite{AK51,Kerker69,BH83}. Special attention is however paid to the shell 
region, where we follow Xu~\cite{X05} and Pe\~{n}a and Pal~\cite{PP09} and split the fields in 
a manner suitable to bring out the hybridization scenario. 
%

The expansions for the incident fields read~\cite{BH83} 
\begin{align}
\vec{E}_{\rm p}^m &=\sum_{n=1}^\infty E_n \big( \vec{M}^{(1)}_{o1n} - i \vec{N}^{(1)}_{e1n} \big)~,\label{Ein} \\
\vec{H}_{\rm p}^m &=-\frac{k_m c}{\omega\mu_m}\sum_{n=1}^\infty E_n \big( \vec{M}^{(1)}_{e1n} \label{Hin}
            + i \vec{N}^{(1)}_{o1n} \big)~,
\end{align}
with expansion coefficients
\begin{align}
E_n=i^nE_0\frac{2n+1}{n(n+1)}~,
\end{align}
where $E_0$ is the strength of the incident electric field.
The radial dependence of these fields is described by Bessel functions of the first kind, indicated by the 
superscript (1), thus they can be interpreted as the penetrating fields (explaining the subscript p) in the 
surrounding medium ($i=m$). The scattered fields outside the particle
may be written as
\begin{align}
\vec{E}_{\rm s}^m &= \sum_{n=1}^\infty E_n \big( ia_n^m\vec{N}^{(3)}_{e1n} - b_n^m \vec{M}^{(3)}_{o1n} \big)~,\label{Es}\\
\vec{H}_{\rm s}^m &= \frac{k_m c}{\omega\mu_m}\sum_{n=1}^\infty E_n \big( ib_n^m\vec{N}^{(3)}_{o1n} \label{Hs}
+ a_n^m \vec{M}^{(3)}_{e1n} \big)~.
\end{align}
The superscript $(3)$ indicates, that Hankel functions of the first kind
$h_n= j_n + \mathrm{i}y_n$ are used for the radial dependence of the fields, whereby $y_n$ are Bessel functions 
of the second kind, which will be indicated by the superscript $(2)$.
Inside the particle the core and the shell region have to be distinguished. The penetrating fields inside the 
core  ($i=1$) are given by 
\begin{align}
\vec{E}_{\rm p}^1 &= \sum_{n=1}^\infty E_n\big( c_n^1\vec{M}_{o1n}^{(1)} - id_n^1\vec{N}_{e1n}^{(1)} \big)~, 
\label{Ecore}\\
\vec{H}_{\rm p}^1 &= -\frac{k_1 c}{\omega\mu_1}\sum_{n=1}^\infty E_n
\big(d_n^1\vec{M}_{e1n}^{(1)} + ic_n^1\vec{N}_{o1n}^{(1)} \big)~,
\label{Hcore}
\end{align}
and the fields inside the shell may be expressed as
\begin{align}
\vec{E}_{ {\rm shell}} &= \sum_{n=1}^\infty E_n \big( f_n\vec{M}^{(1)}_{o1n} - ig_n \vec{N}^{(1)}_{e1n} \nonumber\\
                                        &+ v_n\vec{M}^{(2)}_{o1n} - iw_n \vec{N}^{(2)}_{e1n} \big)~,\label{Eshell2}\\
\vec{H}_{ {\rm shell} } &= -\frac{k_2 c}{\omega\mu_2}
          \sum_{n=1}^\infty E_n \big( g_n \vec{M}^{(1)}_{e1n} + if_n \vec{N}^{(1)}_{o1n} \nonumber\\
            &+ w_n \vec{M}^{(2)}_{e1n} + iv_n \vec{N}^{(2)}_{o1n} \big)~. \label{Hshell2}
\end{align}

The expansions~\eqref{Eshell2} and~\eqref{Hshell2} going back to Aden and Kerker~\cite{AK51} are 
the basis of most applications of the Mie theory to spherical core-shell particles. In our previous 
work we also used them~\cite{THB14,TBH15}. Their physical content however is not obvious. To 
bring the physics to the forefront it is more appropriate to follow the hybridization 
scenario~\cite{Prodan03,PN04,Preston11} and to express the fields inside the shell ($i=2$) as follows~\cite{X05,PP09}
\begin{align}
\vec{E}_{ {\rm shell} } = \vec{E}_{\rm s}^2 + \vec{E}_{ {\rm p} }^2 ~~ \mathrm{and}~~ \vec{H}_{ {\rm shell} } = \vec{H}_{{\rm s}}^2 + \vec{H}_{{\rm p}}^2~,
\label{Eshell1}
\end{align}
where $\vec{E}_{\rm s}^2, \vec{H}_{\rm s}^2$ and $\vec{E}_{{\rm p}}^2, \vec{H}_{{\rm p}}^2$ are, respectively, 
the scattered fields (explaining the subscript $s$) of a cavity consisting of core material 
and embedded in an homogeneous domain of shell material and the penetrating fields of an 
homogeneous particle made out of shell material (hence the subscript $p$) embedded in the 
medium. Adopting Eqs.~\eqref{Es}--\eqref{Hs} and Eqs.~\eqref{Ecore}--\eqref{Hcore} to the 
shell region,
\begin{align}
\vec{E}_{\rm s}^2 &= \sum_{n=1}^\infty E_n \big( ia_n^2 \vec{N}^{(3)}_{e1n} - b_n^2\vec{M}^{(3)}_{o1n} \big)~,\label{Eprop}\\
\vec{H}_{\rm s}^2 &= \frac{k_2 c}{\omega\mu_2}
          \sum_{n=1}^\infty E_n \big( ib_n^2 \vec{N}^{(3)}_{o1n} + a_n^2 \vec{M}^{(3)}_{e1n} \big)~ \label{Hprop}
\end{align}
and 
\begin{align}
\vec{E}_{\rm p}^2 &= \sum_{n=1}^\infty E_n \big( c_n^2\vec{M}^{(1)}_{o1n} - id_n^2 \vec{N}^{(1)}_{e1n} \big)~,\label{Eshell}\\
\vec{H}_{\rm p}^2 &= -\frac{k_2 c}{\omega\mu_2}
          \sum_{n=1}^\infty E_n \big( d_n^2 \vec{M}^{(1)}_{e1n} + ic_n^2 \vec{N}^{(1)}_{o1n} \big)~. \label{Hshell}
\end{align}

Since the total number of expansion coefficients is unchanged, remaining in total eight, 
$a_n^m , b_n^m, a_n^2, b_n^2, c_n^2, d_n^2, c_n^1$, and $d_n^1$, the shell-medium 
boundary conditions at $r=r_2$, 
\begin{align}
\vec{r} \times (\vec{E}_{\rm p}^m + \vec{E}_{\rm s}^m - \vec{E}_{\rm s}^2 - \vec{E}_{\rm p}^2 )&=0~,\label{Eboundary_sm}  \\ 
\vec{r} \times (\vec{H}_{\rm p}^m + \vec{H}_{\rm s}^m - \vec{H}_{\rm s}^2 - \vec{H}_{\rm p}^2 )&=0~, \label{Hboundary_sm}
\end{align}
and the core-shell boundary conditions at $r=r_1$,
\begin{align}
\vec{r} \times (\vec{E}_{\rm s}^2 + \vec{E}_{\rm p}^2 - \vec{E}_{\rm p}^1)&=0~, \label{Eboundary_sc}\\ 
\vec{r} \times (\vec{H}_{\rm s}^2 + \vec{H}_{\rm p}^2 - \vec{H}_{\rm p}^1)&=0~,\label{Hboundary_sc}
\end{align}
are sufficient for their determination. For the derivation of the hybridization scenario we 
focus on the coefficients $d_n^2$ and $a_n^2$ which are resonant for small particles up to 
fourth multipole order (as are all coefficients in front of the spherical vector harmonics
$\vec{N}_{e1}$). 

The hybridization scenario becomes apparent by writing the expansion coefficients $d_n^2$ and 
$a_n^2$ in a form resembling the diagonal elements of a matrix resolvent
\begin{align}
 G(E)=
 \begin{pmatrix}
 \varepsilon_A -E & V \\
 V^*& \varepsilon_B-E
 \end{pmatrix}^{-1}
 \;~
\end{align}
describing the hybridization of two energy levels $\varepsilon_A$ and $\varepsilon_B$. The 
basis for this matrix are the states $|A\rangle$ and $|B\rangle$. Lets consider
\begin{align}
G_A(E) &=\braket{A|G(E)|A}\nonumber\\
       &=\cfrac{1}{g^{-1}_A(E)-\cfrac{|V|^2}{g^{-1}_B(E)}}~
\label{GF}
\end{align}
as a guide, where $|V|^2$ is the hybridization strength of the two levels and $g_i(E)=(E-\varepsilon_i)^{-1}$ 
is the resolvent of the isolated level $i = A, B$. The pole of $g_i(E)$ gives the energy 
of the noninteracting level $i$, while the two poles of $G_A(E)$ are the excitation energies of the 
interacting system. Once the coefficients are in this form their interpretation in terms of an 
hybridization scenario is thus obvious. 

Indeed the coefficients $d_n^2$ and $a_n^2$ obtained from
the boundary conditions~\eqref{Eboundary_sm}--\eqref{Hboundary_sc} can be written as 
\begin{align}
d_n^{2}=\cfrac{X_n^{2}(x_2)}{E_n^{2}(x_2) - \cfrac{X_n^{ { \rm cp}1}(x_1)Z_n^{2}(x_2)}{E_n^{1}(x_1)}} \label{dns}
\end{align}
and
\begin{align}
a_n^{2}=\cfrac{X_n^{ {\rm cp}1}(x_1)X_n^{2}(x_2)/E_n^{2}(x_2)}{E_n^{1}(x_1) - 
\cfrac{X_n^{ { \rm cp} 1}(x_1) Z_n^{2}(x_2)}{E_n^{2}(x_2)}}~  \label{anp}
\end{align}
with $E_n^{1}$, $E_n^{2}$, and $X_n^{{\rm cp}1}Z_n^{2}$ playing the role of $g_A^{-1}$, 
$g_B^{-1}$, and $|V|^2$, respectively. The function $X_n^{2}$ has no direct analogue. For better readability 
we postpone the definitions of these functions to the point where they enter the discussion of the 
physical content of Eqs.~\eqref{dns} and~\eqref{anp}. 

A comparison of~\eqref{dns} and~\eqref{anp} with~\eqref{GF} suggests, now at the level of 
the Mie coefficients, that the optical response of a core-shell particle is the response of 
two hybridized subsystems, with individual resonances determined by 
\begin{align}
E_n^{2}(x_2) &= N_2 \xi^\prime_n(N_m x_2) \psi_n(N_2 x_2)  \nonumber\\
            &- N_m\xi_n(N_m  x_2) \psi^\prime_n(N_2 x_2)=0~,
\label{E_ds}
\end{align}
and
\begin{align}
E_n^{1}(x_1) &= N_1 \xi^\prime_n(N_2x_1) \psi_n(N_1x_1) \nonumber\\
            &- N_2\xi_n(N_2 x_1) \psi^\prime_n(N_1 x_1) =0~,
\label{E_dc}
\end{align}
respectively, and an hybridization strength given by $X_n^{ {\rm cp}1}(x_1)Z_n^{2}(x_2)$. We have chosen all components to be nonmagnetic ($\mu_{1,2,m}=1$). 
Notice in Eqs.~\eqref{E_ds} and~\eqref{E_dc} the Ricatti-Bessel functions $\psi_n(\rho)=\rho j_n(\rho)$ 
and $\xi_n(\rho)=\rho h_n(\rho)$ as well as the implicit definitions of the functions $E_n^{2}$ 
and $E_n^{1}$ appearing in~\eqref{dns} and \eqref{anp}. The physical content 
of~\eqref{E_ds} and \eqref{E_dc} can be deduced by comparing these equations with the 
denominator of the scattering coefficient describing an homogeneous particle~\cite{BH83}
\begin{align}
a_n^m=\frac{N_1\psi^\prime_n(N_mx)\psi_n(N_1 x)-N_m \psi_n(N_mx) \psi^\prime_n(N_1x) }{N_1\xi_n^\prime(N_mx)\psi_n(N_1x) 
	- N_m \xi_n(N_mx) \psi^\prime_n(N_1x) }~,
\label{a_ncomp}
\end{align}
which has to vanish in the resonance case. Here, $N_1$ and $N_m$ are the refractive indices of the particle and the surrounding medium,
respectively, and $x$ is the size parameter of the particle. Clearly, Eq.~\eqref{E_ds} 
is the resonance condition of an homogeneous particle with size parameter $x_2$ and refractive 
index $N_2$ embedded in a medium specified by $N_m$, while Eq.~\eqref{E_dc} is the resonance 
condition of an homogeneous particle with size parameter $x_1$ and refractive index $N_1$ embedded 
in a medium characterized by $N_2$. If the real part of the dielectric function of the core
$\varepsilon'_1>0$ and the real part of the dielectric function of the shell $\varepsilon'_2<0$ 
it is however the embedding medium which supports the surface modes. It is thus appropriate to 
consider~\eqref{E_dc} as the resonance condition of a cavity filled with material described by $N_1$
and embedded in a medium characterized by $N_2$. 

So far we extracted from the Mie formulae the two elementary building blocks, a shell-embedded 
core cavity and a medium-embedded homogeneous shell particle, whose surface modes hybridize to make up the optical 
response of the core-shell particle. We now turn to the hybridization strength $X_n^{ {\rm cp}1}(x_1)Z_n^{2}(x_2)$. 
It is interesting to analyze under what conditions it vanishes and what this implies for the expansion 
coefficients. The two subsystems are noninteracting when at least one of the following conditions holds:  
\begin{align}
X^{ {\rm cp} 1}_n(x_1) &= N_1 \psi_n(N_2x_1) \psi^\prime_n(N_1 x_1)  \nonumber\\
            &- N_2 \psi^\prime_n(N_2x_1) \psi_n(N_1x_1)=0~,\label{Xapvanish}\\ 
Z^{2}_n(x_2) &= N_2 \xi_n^\prime(N_m x_2) \xi_n(N_2 x_2) \nonumber\\
            &- N_m \xi_n(N_m x_2) \xi_n^\prime(N_2 x_2)=0~.
\label{Zadvanish}
\end{align}

Condition~\eqref{Xapvanish}, which implicitly defines $X_n^{ {\rm cp}1}$, is satisfied when the refractive 
indices of the core and the shell are the same, $N_1=N_2$. It leads to $a_n^2=0$ and 
\begin{align}
      d_n^{2} &= \frac{X_n^{2}(x_2)}{E_n^{2}(x_2)} \nonumber\\
              &= \frac{N_2 \xi^\prime_n(N_m x_2) \psi_n(N_m x_2) - N_2\xi_n(N_m x_2) \psi^\prime_n(N_m x_2) }{N_2 
                 \xi^\prime_n(N_m x_2) \psi_n(N_2 x_2) - N_m\xi_n(N_m x_2) \psi^\prime_n(N_2 x_2)} \nonumber \label{dn_homo}\\
              &= d_n^1~,
\end{align}
where $d_n^1$ is the coefficient arising in the expansions~\eqref{Ecore} and~\eqref{Hcore} 
for the penetrating fields inside the core which is now identical with the shell. Note, the 
numerator in the second equality defines the function $X_n^{2}$ 
appearing in Eqs.~\eqref{dns} and~\eqref{anp}. Thus, due to the absence of the core-shell 
interface, the scattering fields in the shell region vanish, leading to $a_n^2=0$, and the penetrating  fields inside the shell become equivalent to the penetrating fields of the core, signalled by $d_n^2=d_n^1$, with expansion 
coefficients describing the fields inside an homogeneous particle with refractive index $N_2$ 
embedded in a medium with refractive index $N_m$. The core-shell particle is in this 
limit reduced to a medium-embedded homogeneous shell particle. 

The second condition~\eqref{Zadvanish}, which implicitly defines $Z_n^{2}$, reduces the 
core-shell particle to a cavity embedded in an homogeneous medium. In this case the 
refractive indices of the medium and the shell have to be identical, $N_m=N_2$, leading to 
$d_n^{2}=1$ and
\begin{align}
a_n^2 =\frac{N_1 \psi^\prime_n(N_2 x_1) \psi_n(N_1 x_1) - N_2 \psi_n(N_2 x_1) \psi^\prime_n(N_1 x_1) 
}{N_1 \xi^\prime_n(N_2 x_1) \psi_n(N_1 x_1) - N_2 \xi_n(N_2 x_1) \psi^\prime_n(N_1 x_1) } \label{ap}~.
\end{align}
Hence, due to the missing medium-shell interface, the penetrating fields of the shell become equal  
to the incoming fields, which are considered to be the penetrating fields in the medium, as can be seen by comparing Eqs.~\eqref{Eshell} and~\eqref{Hshell} 
for $d_n^2=1$ with Eqs.~\eqref{Ein} and~\eqref{Hin}. The expansion coefficient $a_n^2$ 
is in this limit attached to the scattered fields of a cavity. Indeed, looking at 
the scattering coefficient of an homogeneous sphere~\cite{BH83} $a_n^m$ given by Eq.~\eqref{a_ncomp} 
and substituting $N_m\to N_2$ and $x\to x_1$ makes~\eqref{a_ncomp} 
identical to \eqref{ap}. Hence, for $N_m=N_2$ the core-shell particle is reduced to a 
core cavity embedded in a shell medium. 

It should be noted that in the standard expansion the coefficients $f_n, g_n, v_n$, and $w_n$
reduce in the respective limits also to the Mie coefficients of a medium-embedded homogeneous 
shell particle and a shell-embedded core-cavity~\cite{TBH15}. What we have shown by working
with $a_n^2$, $b_n^2$, $c_n^2$, and $d_n^2$ and rewriting them in a particular manner is that 
the two limiting cases are actually two building blocks whose resonances are always virtually 
present. They are encoded in the functions $E_n^{2}(x_2)$ and $E_n^{1}(x_1)$. The interaction 
$X_n^{ {\rm cp }1}(x_1)Z_n^{2}(x_2)$, controlled by the geometry and the material parameters, 
defines their lifetimes and makes the optical response of the core-shell particle given 
by the poles of $d_n^2$ and $a_n^2$ unique, in full accordance with the hybridization 
scenario~\cite{Prodan03,PN04,Preston11}. 
\begin{table}[b]
\begin{center}
  \begin{tabular}{c|c|c|c|c}
    model  & $\lambda^{-1}_{\rm TO}$ (${\rm cm}^{-1}$) & ~~$\varepsilon_0$~~ & ~~$\varepsilon_\infty$~~ \\\hline
    core   & 300 & 3  & 2 \\
    shell  & 600 & 20 & 2 \\
  \end{tabular}
  \caption{Parameters for the model core-shell particle.}
  \label{Parameters}
\end{center}
\end{table}

Before presenting numerical results for finite hybridization strength let us consider the case 
where the hybridization is turned off by a vanishing filling factor $f=r_1/r_2=x_1/x_2$. Depending on 
how the limit $f\to 0$ is taken, the core-shell particle reduces again to either one of its 
building blocks, a medium-embedded homogeneous shell particle or a shell-embedded core-cavity. 
If the filling factor vanishes because the core becomes smaller and smaller, that is, 
because $x_1 \to 0$, while $x_2$ is fixed, it is $X^{ {\rm cp }1}_n(x_1)$ which vanishes. As discussed above, 
the core-shell particle is then reduced to an homogeneous particle made out of shell material 
and embedded in a medium characterized by a refractive index $N_m$. If the filling factor 
vanishes however because the particle becomes larger and larger, while the core size is 
fixed, that is, for $x_1$ fixed and $x_2 \to \infty$, it is $Z^{2}_n(x_2)$ which vanishes and 
reduces, according to the discussion given in the previous paragraph, the core-shell particle 
to a cavity made out of core material embedded in an homogeneous surrounding made out of shell 
material. For finite filling factors both $X^{ {\rm cp }1}_n(x_1)$ and $Z^{2}_n(x_2)$ are finite and 
mix the optical response of the building blocks. This will be discussed in the next section.
\begin{figure}[t]
\includegraphics[width=0.9\linewidth]{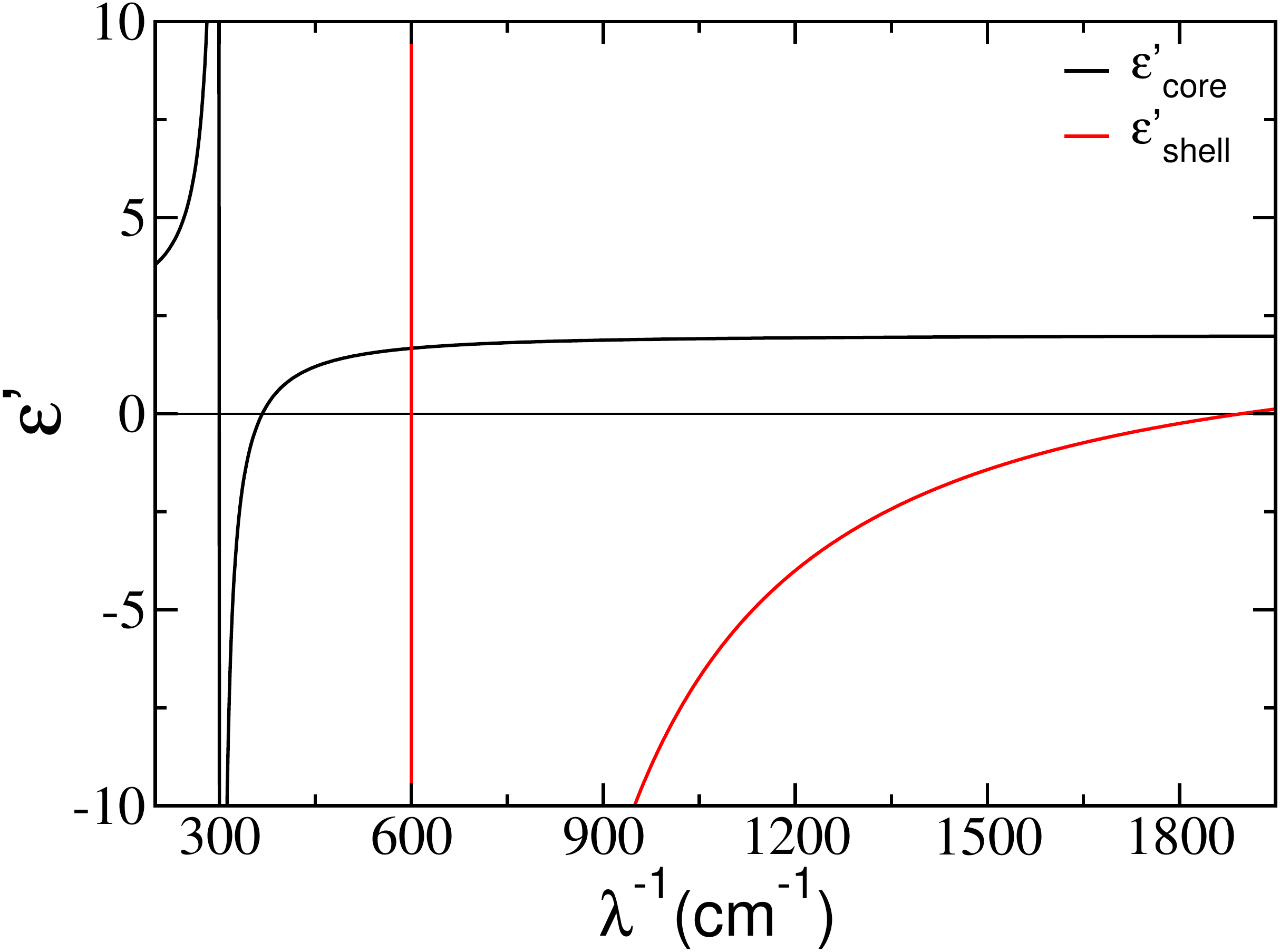}%
\caption{\label{fig2} (Color online) Real parts of the dielectric functions specified 
by Eq.~\eqref{EpsModel} and the parameters in Table~\ref{Parameters} as a function of the 
wave number $\lambda^{-1}$.} 
\end{figure}

The re-organization of the Mie theory discussed in this section for the particular case 
of a spherical core-shell particle can be applied to other composite particles as well
provided the symmetry is high enough to yield analytical expressions for the expansion 
coefficients of the electromagnetic fields. In the Appendix we show this for a 
stratified sphere~\cite{W62,B85,SQ94,X05}. 


\section{Illustration}\label{Results}

To illustrate the classical optical response of a spherical core-shell particle we consider a 
particle with a dielectric core and a dielectric shell. Previously we proposed to use this type of 
particle in a gas discharge as an electric probe with optical read-out~\cite{THB14}. The idea, 
formulated initially for an homogeneous dielectric particle~\cite{HBF12b}, is to determine 
the electric field at the particle's position in the discharge from the balance of the forces 
acting on it and the charge-dependent shift of one of its extinction resonances. The shifts can 
be maximized by localizing inside the shell the elementary charges the particle acquired from 
the plasma using materials with negative electron affinity as a core and materials with 
positive electron affinity as a shell.

The extinction spectra we calculated with the goal of employing them as a charge diagnostics 
in a plasma are based on the complex dielectric functions of the real materials~\cite{THB14,HBF12b}. 
In our work on topological aspects of light scattering by dielectric core-shell particles~\cite{TBH15}
we found however dissipation to blur higher order bonding and antibonding resonances. For  
the present purpose we employ therefore dissipationless model dielectric functions with 
parameters chosen such that bonding and antibonding resonances can be clearly identified
up to the third multipole order. 
\begin{figure}[t]
\includegraphics[width=\linewidth]{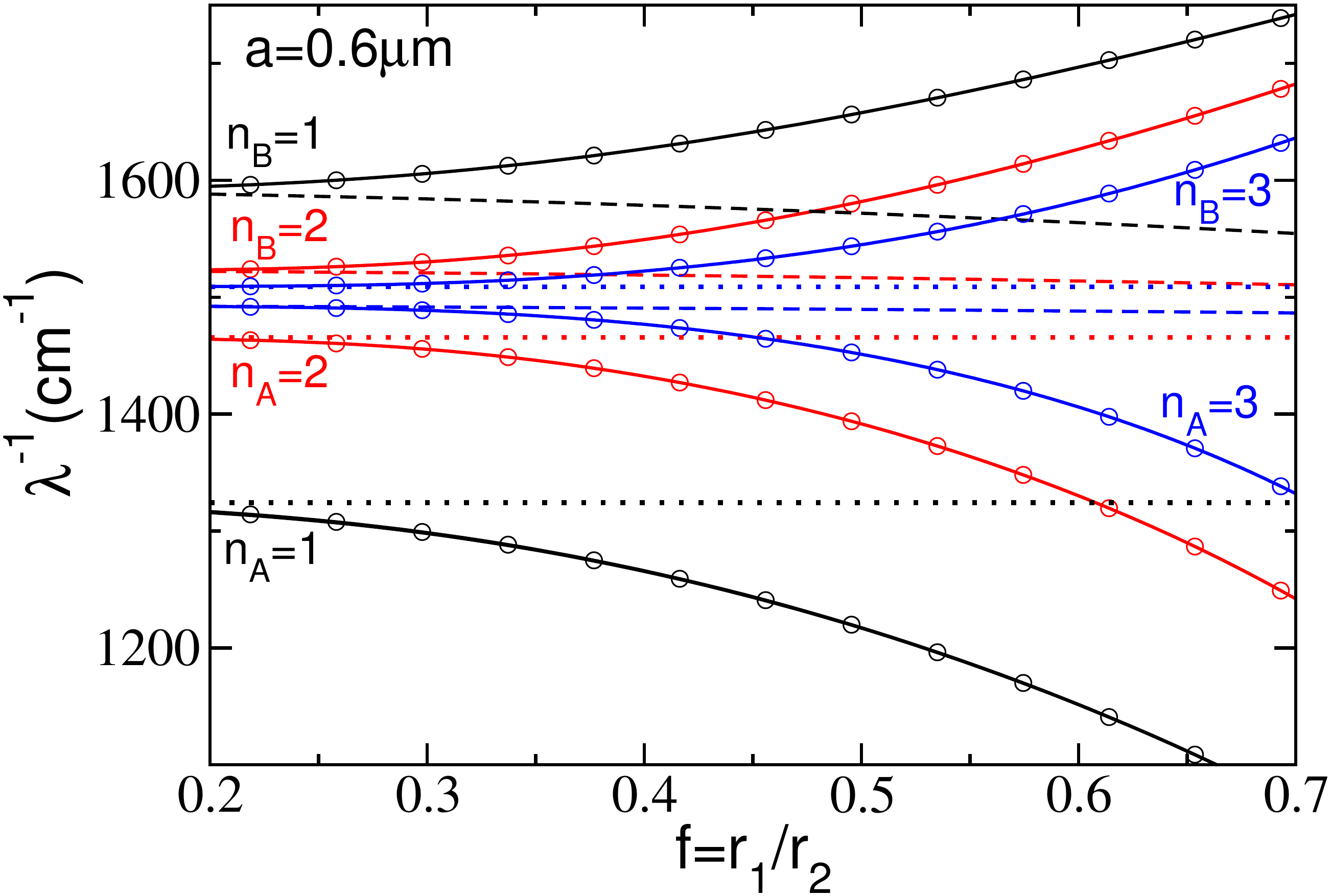}%
\caption{\label{fig3} (Color online) Solid lines show the wave numbers of the bonding
(subscript A) and antibonding (subscript B) dipole ($n_{\rm A,B}=1$), quadrupole ($n_{\rm A,B}=2$),
and hexapole ($n_{\rm A,B}=3)$ resonance as a function of the filling factor $f=r_1/r_2$ for
a core-shell particle with radius $r_2=0.6\,\mu$m. The dotted and dashed lines denote,
respectively, the positions of the resonances of the medium-embedded homogeneous shell
particle and the shell-embedded core cavity, while bullets indicate the positions of the
maxima in the extinction spectrum.}
\end{figure}

As in our work on topological aspects of light 
scattering by dielectric core-shell particles~\cite{TBH15}, we set $N_m=1$, that is, embed
the particle in vacuum, and use
\begin{align}
\varepsilon^\prime = \varepsilon_{\infty} + \lambda^{-2}_{\rm TO}\frac{\varepsilon_0 - \varepsilon_{\infty}}
                     {\lambda^{-2}_{\rm TO} - \lambda^{-2}}~,~~\varepsilon^{\prime\prime} = 0~,
\label{EpsModel}
\end{align}
with $\lambda_{\rm TO}^{-1}$, the wave number of the transverse optical phonon, and $\varepsilon_0$ 
and $\varepsilon_\infty$, the dielectric constants at large and small wave numbers, given in 
Table~\ref{Parameters} for the core and the shell, respectively. 
The real parts of the model dielectric functions are shown in Fig.~\ref{fig2}.
Of particular interest is the spectral range where surface modes at the core-shell interface
are excited, that is, the range of wave numbers where the real part of the dielectric function 
of the shell is negative and the real part of the dielectric function of the core is positive. 
For the model dielectric functions plotted in Fig.~\ref{fig2} this is the case for 
$600\,{\rm cm}^{-1} < \lambda^{-1} < 1900\,{\rm cm}^{-1}$.  

The resonances in this spectral range are depicted in Fig.~\ref{fig3} as a function of the filling 
factor $f=r_1/r_2$ for a particle with radius $r_2=0.6\,\mu$m. In accordance with the
hybridization scenario~\cite{Prodan03,PN04,Preston11} and our previous work~\cite{TBH15,THB14}
we label the resonances at lower wave numbers bonding (subscript A) and the resonances at higher 
wave numbers antibonding (subscript B). Justification for this labeling comes from the 
polarities of the induced surface charges at the two interfaces which are in-phase for the 
bonding and out-of-phase for the antibonding resonances~\cite{TBH15}. The solid lines give 
the positions for the bonding and antibonding dipole ($n_{\rm A,B}=1$), quadrupole ($n_{\rm A,B}=2$), 
and hexapole ($n_{\rm A,B}=3$) resonance as obtained from the poles of Eq.~\eqref{dns}, while the 
dotted and dashed lines give, respectively, the solutions of Eqs.~\eqref{E_ds} and~\eqref{E_dc}, 
that is, the positions of the resonances of the medium-embedded homogeneous shell particle and the 
shell-embedded core cavity.  Also shown are the positions of the maxima in the extinction efficiency, 
$Q_t=\frac{2}{x_2}\sum_{n=1}^\infty (2n+1) \mathrm{Re}[a_n^m + b_n^m]$, where $a_n^m$ and $b_n^m$ are the
expansion coefficients of the scattered fields $\vec{E}_{\rm s}^m$ and $\vec{H}_{\rm s}^m$ (see Eqs.~\eqref{Es} 
and~\eqref{Hs}). The resonances in the extinction spectrum depicted by the bullets follow precisely the 
resonances in the expansion coefficients $d_n^2$ and $a_n^2$ indicating that the denominator of  
$a_n^m$, which is the resonant coefficient in the expansion of the scattered fields, can be written in 
the same form as the denominators of $d_n^2$ and $a_n^2$.
\begin{figure}[t]
\includegraphics[width=\linewidth]{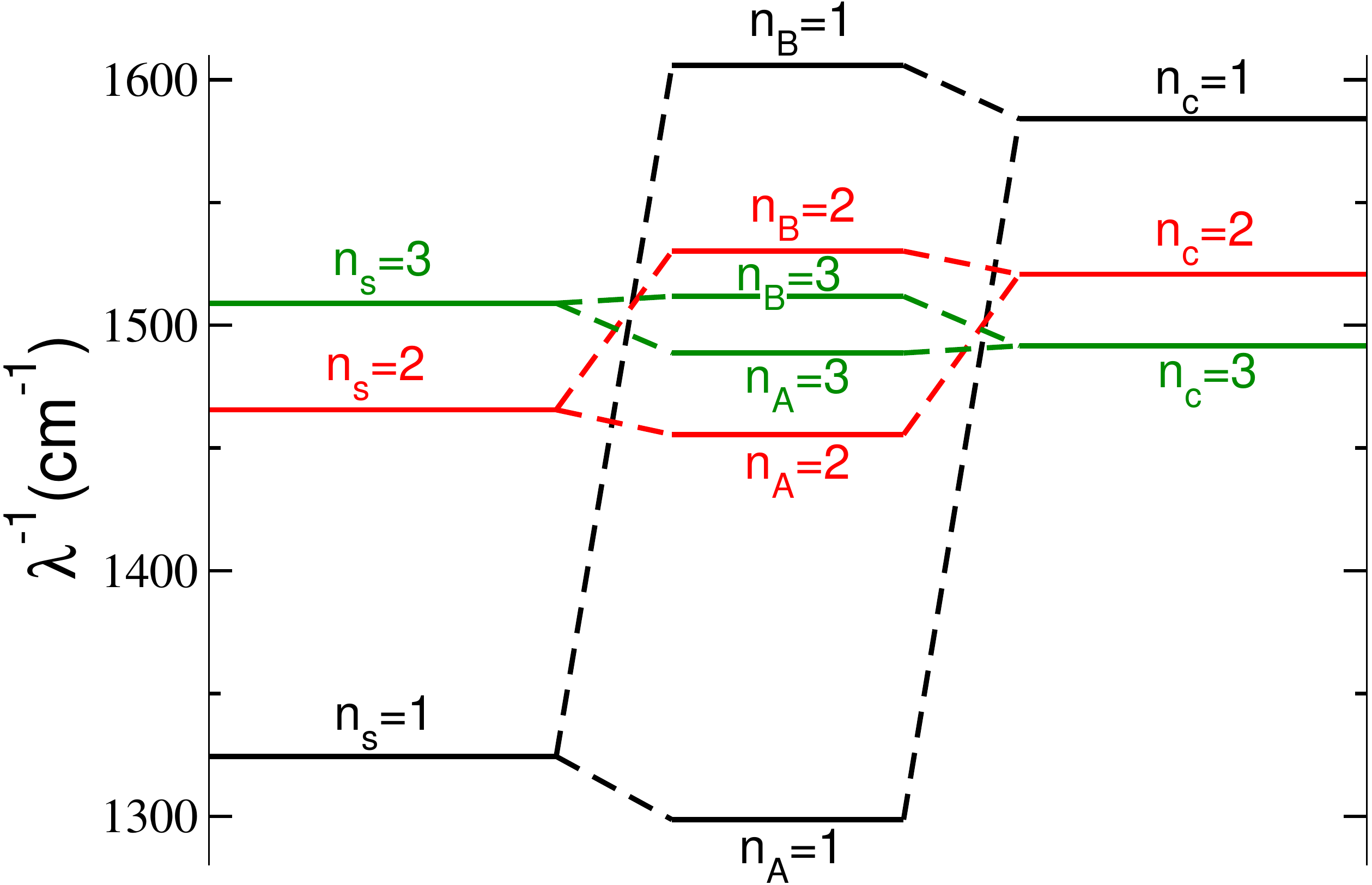}%
\caption{\label{fig4} (Color online) Energetic positions of the resonances 
of the medium-embedded homogeneous shell particle ($n_s=1,2,3$), the shell-embedded core cavity 
($n_c=1,2,3)$, and the core-shell particle ($n_{A,B}=1,2,3$) for $f=0.3$ and 
$r_2=0.6\,\mu$m. From the decreasing deviations of the hybridized from the non-interacting 
resonances can be inferred that the hybridization strength $X_n^{{\rm cp}1}(x_1)Z_n^{2}(x_2)$ 
decreases with increasing multipole order.
}
\end{figure}

The main feature of Fig.~\ref{fig3} is the increasing splitting between the bonding 
and antibonding resonances with increasing filling factor $f=r_1/r_2$. As expected from
the hybridization scenario the hybridization strength $X_n^{ {\rm cp }1}(x_1)Z_n^{2}(x_2)$
controlling the splitting increases with $f$ because the decrease in shell 
thickness leads to a stronger coupling between the surface modes at the core-shell 
and the shell-medium interface. For $f\approx 0.2$ the 
resonances of the core-shell particle merge with the resonances of its building 
blocks, implying that the hybridization strength becomes vanishingly small. Notice, 
for $n_{\rm A,B}\le 2$ the bonding resonances merge with the resonances of the 
medium-embedded homogeneous shell particle, that is, the solutions of $E_n^{2}(x_2)=0$, 
while the antibonding resonances merge with the resonances of the shell-embedded core 
cavity, that is, the solutions of $E_n^{1}(x_1)=0$. For $n_{\rm A,B}=3$, however, it is 
the other way around. The reason is simply the energetic ordering of the resonances 
of the building blocks. For $n_{\rm A,B}\le 2$ the resonances of the shell-embedded
core cavity are energetically above the resonances of the medium-embedded homogeneous 
shell particle, while for $n_{\rm A,B}=3$ the ordering is reversed. Figure~\ref{fig4} 
displays this ordering for $f=0.3$. 

\begin{figure*}
\includegraphics[width=\linewidth]{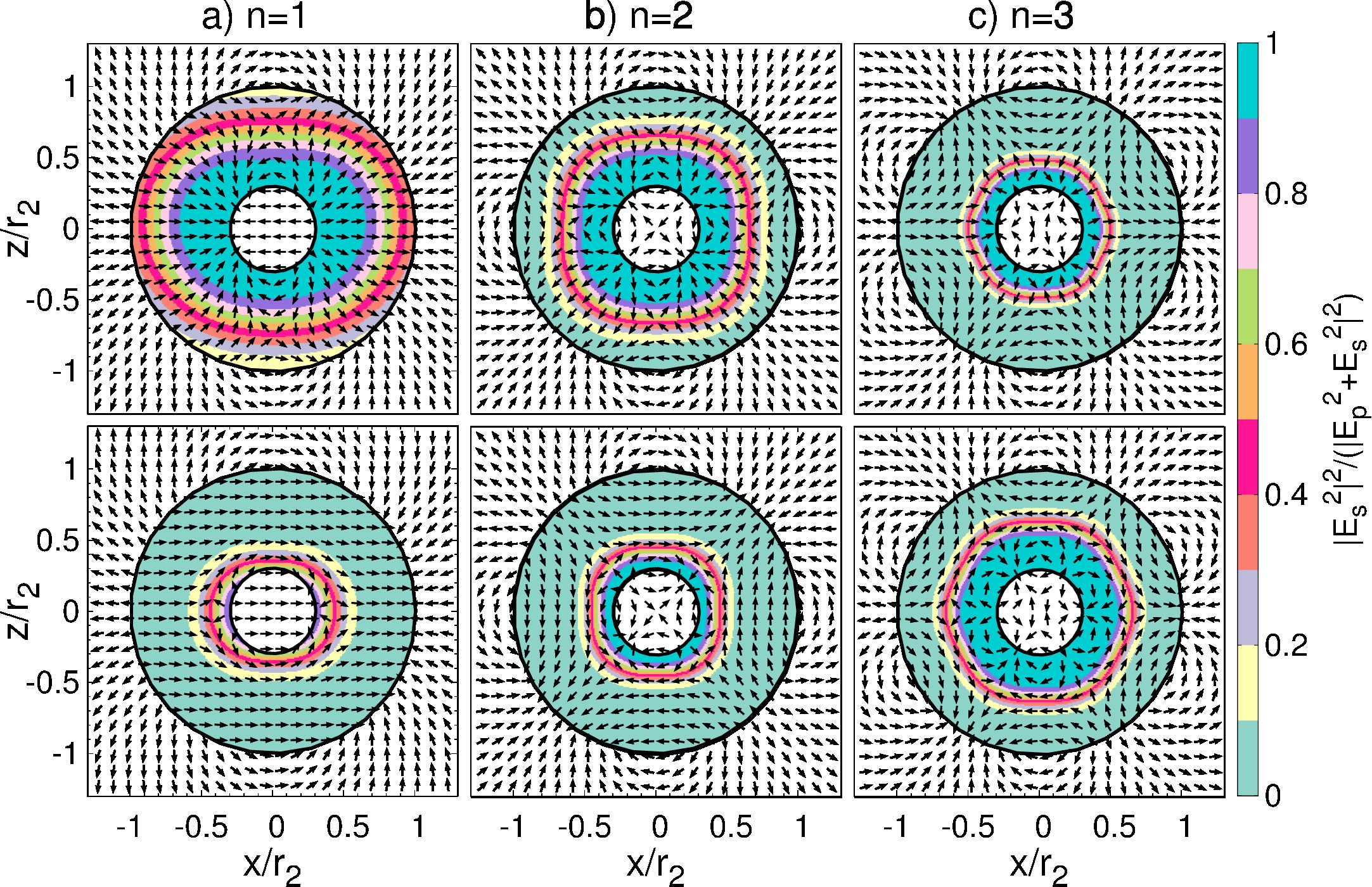}%
\caption{\label{fig5} (Color online) Electric field distributions in the $xz$ plane of a core-shell particle
($r_2=0.6\,\mu$m, $f=0.3$) near the bonding (lower panels) and antibonding (upper panels) dipole ($n=1$), 
quadrupole ($n=2$), and hexapole ($n=3$) resonances. The wavenumbers are $\lambda^{-1}=1313.62\,$cm$^{-1}$ and
$\lambda^{-1}=1607.85\,$cm$^{-1}$ for the bonding and antibonding dipole, $\lambda^{-1}=1455.5\,$cm$^{-1}$
and $\lambda^{-1}=1530.22\,$cm$^{-1}$ for the bonding and antibonding quadrupole, and
$\lambda^{-1}=1488.706\,$cm$^{-1}$ and $\lambda^{-1}=1511.805\,$cm$^{-1}$ for the bonding and antibonding
hexapole resonance. The direction of the total electric field is shown by black arrows and the color
coding gives the intensity of the scattering fields in units of the intensity of the overall electric
fields inside the shell. Black circles indicate the particle and its core.}
\end{figure*}

Another way to visualize the hybridization scenario for the core-shell particle 
is to look at the electric fields inside the shell. This is shown in Fig.~\ref{fig5}
for the dipole ($n=1$), quadrupole ($n=2$), and hexapole ($n=3$) resonances of a 
particle with radius $r_2=0.6\,\mu$m and filling factor $f=0.3$.
The lower and upper panels in Fig.~\ref{fig5} depict respectively the bonding 
and antibonding resonances. That the assignment is correct can be inferred from the black arrows 
visualizing the orientation of the electric field inside and outside the particle. From them 
the polarity of the induced surface charges follows verifying that it is in-phase for the 
bonding and out-of-phase for the antibonding resonances. The wave numbers are tuned to maximize
the spatial extension of the scattered fields outside the particle, $\vec{E}_{\rm s}^m$ and 
$\vec{H}_{\rm s}^m$, taking as a measure the distance of the singular points~\cite{WL04,TL06} in 
the outer Poynting field from the center of the particle (for a discussion of singular points 
in the dipole fields of dielectric core-shell particles see Ref.~\cite{TBH15}). From the ratio of 
the intensity of the scattering electric field inside the shell $|\vec{E}_{\rm s}^2|^2$ to the intensity of 
the overall electric field inside the shell $|\vec{E}_{ \rm p}^2+\vec{E}_{ \rm s}^2|^2$, shown by the color 
coding, can be moreover deduced that the influence of the scattering fields is for the antibonding hexapole 
resonance smaller than for the bonding one, in contrast to the dipole and quadrupole resonances where it 
is reversed. The reason is again the energetic ordering of the resonances plotted in Fig.~\ref{fig4}. 
The hexapole resonance of the shell-embedded core cavity drops below the hexapole resonance of the 
medium-embedded homogeneous shell particle. As a result, for $n=3$ it is the bonding resonance which 
acquires more cavity character, and is thus more affected by the scattering fields, and not the antibonding 
one as it is the case for $n=1,2$. 

For core-shell particles with dissipation, having complex dielectric functions, the higher order bonding 
and antibonding resonances may be blurred. For a \CaO/\AlTwoOThree\ particle, for instance, only the 
dipole and quadrupole resonances can be clearly identified~\cite{TBH15}. Hence, higher order bonding and 
antibonding resonances may be distinguishable only for a judicious choice of materials. The multipole
where the character of the bonding and antibonding resonance changes depends also on the materials. 
It can occur already for the dipole~\cite{BG10}. The character of the resonances determines
the number and locations of the singular points as well as the spatial distribution of the 
dissipation inside the core-shell particle~\cite{TBH15}. Changing it by tuning the resonances 
of its building blocks may thus be interesting from a physics and an application point of view.\\

\section{Conclusions} \label{Conclusions}

The hybridization scenario for the classical optical response of composite objects states 
that it can be understood in terms of interacting surface modes arising at the interfaces 
of the object. 

For the particular case of a spherical core-shell particle we derived the 
scenario directly from the expansion of the electromagnetic fields inside and outside the 
particle in terms of vector spherical harmonics. By re-organizing the fields inside the 
shell region into the penetrating fields of a shell particle and the scattered fields of 
a core cavity we were able to derive formulae for the expansion coefficients resembling 
the diagonal elements of a matrix resolvent describing two hybridized energy levels. From the 
formulae the building blocks of the hybridization scenario, the shell-embedded core cavity 
and the medium-embedded homogeneous shell particle, as well as the coupling between them could be 
straightforwardly identified. The physical content of the coefficients became thus immediately
clear. In an Appendix we also demonstrated that the same strategy can be applied to a stratified 
sphere containing an arbitrary number of shells. Re-organizing the expansion coefficients for the 
fields yields again expressions which resemble the diagonal elements of a matrix resolvent 
describing a chain of hybridized energy levels. The basic building blocks of the hybridization 
scenario and their coupling could thus be identified directly from the expansion coefficients. 

Initially the hybridization scenario has been deduced in the electrostatic approximation 
for small spherical stratified particles without retardation using equations for the surface 
charges induced at the interfaces. The derivation we presented  
includes retardation and is thus applicable to spherical stratified particles of any size. 
In addition it provides a road-map for analytically deriving the hybridization scenario 
for other composite objects with high symmetry for which analytical solutions of the Mie 
theory are available. The physical mechanisms buried in these solutions may thus become 
apparent.

\begin{acknowledgments}
This work was supported by the Deutsche Forschungsgemeinschaft through the Transregional 
Collaborative Research Center SFB/TRR24. 
\end{acknowledgments}

\section*{Appendix}
In this Appendix we re-organize the Mie theory of a stratified spherical 
particle~\cite{W62,B85,SQ94,X05} to obtain the hybridization scenario also for this 
situation.

\begin{figure}[t]
	\includegraphics[width=0.9\linewidth]{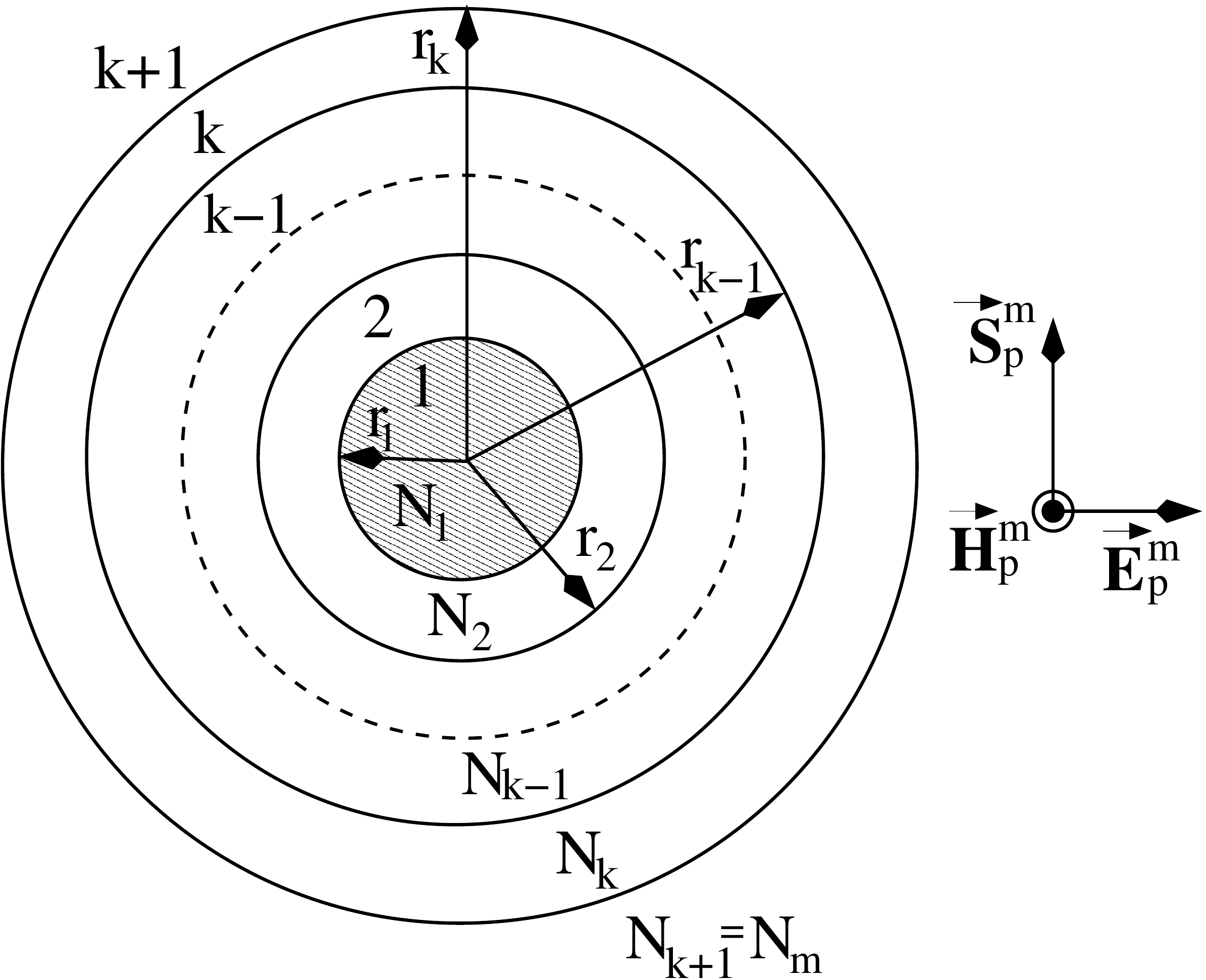}%
	\caption{\label{A_fig1} Geometry of light scattering by a stratified sphere with 
        total radius $r_k$ containing $k$ layers with radii $r_i$ for $1 \leq i \leq k$, 
        whereby the core is counted as the first layer $(i=1)$. Refractive indices $N_i$ 
        characterize each layer and the medium $(i=k+1=m)$. The incident electromagnetic 
        plane wave is described by a Poynting vector $\vec{S}^m_{\rm p}$ propagating in 
        $z$ direction, an electric field $\vec{E}_{\rm p}^m$ polarized in $x$ direction,
        and a magnetic field $\vec{H}_{\rm p}^m$ along the $y$ direction.}
\end{figure}

As indicated in Fig.~\ref{A_fig1} the particle contains $k$ layers. The core is counted as 
the first layer $i=1$, while the outermost layer and the surrounding medium are labelled 
$i=k$ and $i=k+1=m$, respectively. The label $i$ is used for both the $i$th shell and its 
interface to the $(i+1)$th shell. The abbreviations of the main text are adopted. Each 
layer $i$ is characterized by its radius $r_i$, its magnetic permeability $\mu_i$ and its 
refractive index $N_i(\lambda^{-1})=\sqrt{\varepsilon_i(\lambda^{-1})}$, where 
$\varepsilon_{i}(\lambda^{-1})=\varepsilon_{i}^\prime(\lambda^{-1})+i\varepsilon_{i}^{\prime\prime}(\lambda^{-1})$ 
is the complex dielectric function for the respective region. The total radius of the 
particle is the radius of the outermost layer $r_k$. As before we use $k_i=\kappa N_i$ with 
$i=1,...,k,k+1$, where $N_{k+1}=N_m$ is the refractive index of the surrounding medium, and 
the size parameters $x_i=2\pi\lambda^{-1} r_i$.

\begin{figure}[t]
        \includegraphics[width=0.9\linewidth]{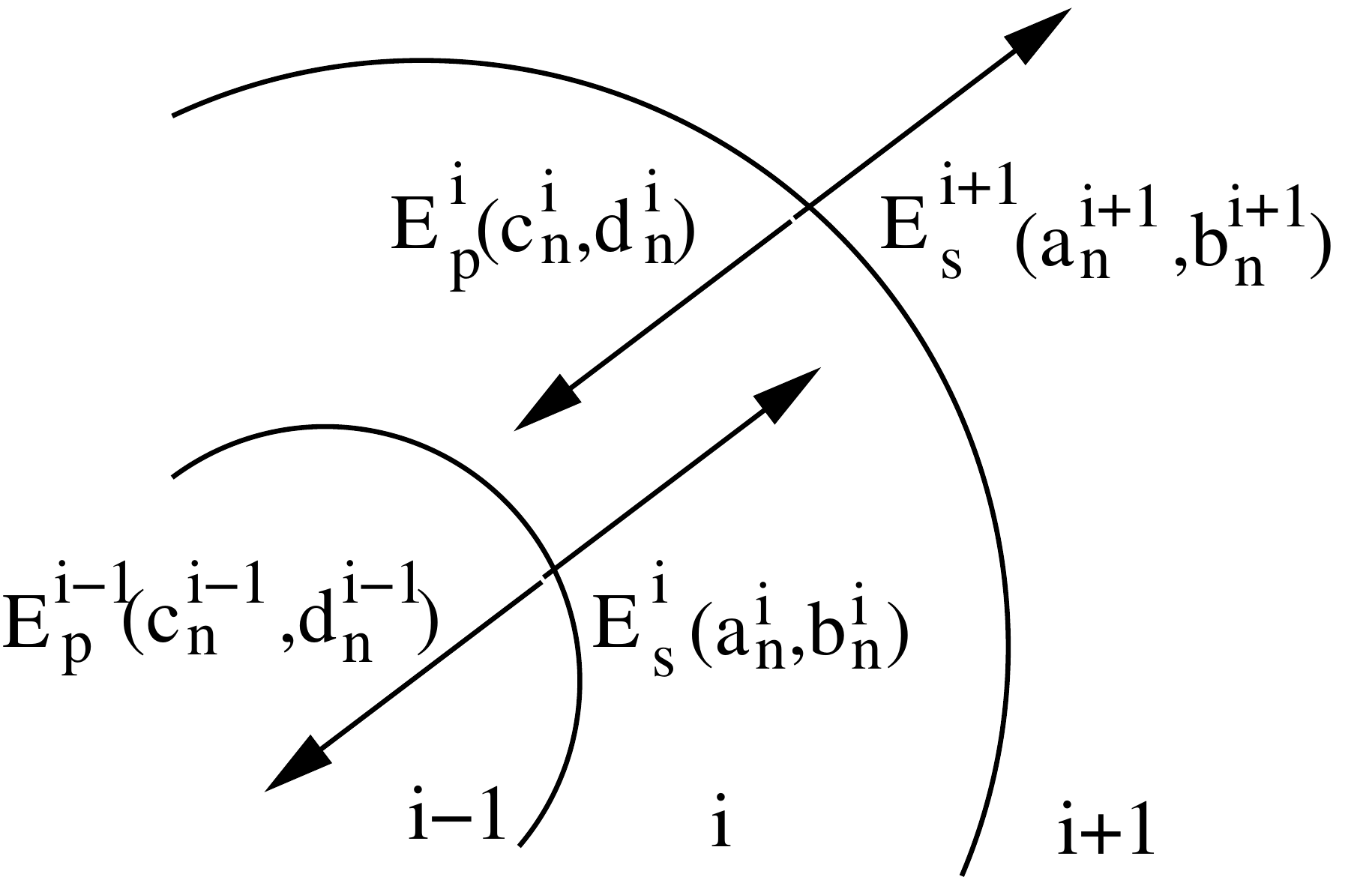}%
        \caption{\label{A_fig2} Superposition of the electric fields in the shell region $i$. The
        resonance arising at the interface $i$ between the $i$th and $(i+1)$th shell is characterized
        by a scattered field $\vec{E}_{\rm s}^{i+1}$ propagating into the $(i+1)$th shell and a
        field $\vec{E}_{\rm p}^i$ penetrating into the $i$th shell. }
\end{figure}

The expansions for the incident and the scattered fields outside the particle as well 
as for the penetrating fields inside the core, given respectively by Eqs.~\eqref{Ein} 
and \eqref{Hin}, \eqref{Es} and \eqref{Hs}, and \eqref{Ecore} and \eqref{Hcore}, remain 
the same, while the splitting of the fields (and their interpretation) in the shell of 
the core-shell particle, viz. Eq.~\eqref{Eshell1}, is now adopted to each shell 
$2\leq i\leq k$ of the stratified sphere. Hence, 
\begin{align}
\vec{E}_{ \rm shell}^i = \vec{E}_{\rm s}^i + \vec{E}_{\rm p}^i ~~ \mathrm{and}~~ \vec{H}_{\rm shell}^i = \vec{H}_{ \rm s}^i + \vec{H}_{ \rm p}^i~,
\label{A_Eshelli}
\end{align}
whereby the scattered fields of the $(i-1)$th subsystem
\begin{align}
\vec{E}_{\rm s}^i &= \sum_{n=1}^\infty E_n \big( ia^{i}_n \vec{N}^{(3)}_{e1n} - b^{i}_n\vec{M}^{(3)}_{o1n} \big)~,\label{Eprop_i}\\
\vec{H}_{\rm s}^i &= \frac{k_i c}{\omega\mu_i}
\sum_{n=1}^\infty E_n \big( ib^{i}_n \vec{N}^{(3)}_{o1n} + a^{i}_n \vec{M}^{(3)}_{e1n} \big)~ \label{Hprop_i} 
\end{align}
interact with the penetrating fields of the $i$th subsystem
\begin{align}
\vec{E}_{\rm p}^i &= \sum_{n=1}^\infty E_n \big( c^{i}_n\vec{M}^{(1)}_{o1n} - id^{i}_n \vec{N}^{(1)}_{e1n} \big)~,\label{Eshell_i}\\
\vec{H}_{\rm p}^i &= -\frac{k_i c}{\omega\mu_i}
\sum_{n=1}^\infty E_n \big( d^{i}_n \vec{M}^{(1)}_{e1n} + ic^{i}_n \vec{N}^{(1)}_{o1n} \big)~, \label{Hshell_i}
\end{align}
as schematically shown Fig.~\ref{A_fig2}. A resonance arising at the $i$th interface, which--in the counting 
scheme introduced above--separates the $i$th from the $(i+1)$th shell, is thus described by a scattered field 
propagating into the $(i+1)$th shell and a penetrating field propagating into the $i$th shell. The overall 
field in the $i$th shell is a superposition of the scattering fields of the $(i-1)$th subsystem and the 
penetrating fields of the $i$th subsystem.

Again the expansion coefficients are determined by the electromagnetic boundary conditions. For the 
shell-medium interface at $r=r_k$, 
\begin{align}
	\vec{r} \times (\vec{E}_{\rm p}^{k+1} + \vec{E}_{\rm s}^{k+1} - \vec{E}_{\rm s}^k - \vec{E}_{\rm p}^k )&=0~,\label{Eboundary_sm_n}  \\ 
	\vec{r} \times (\vec{H}_{\rm p}^{k+1} + \vec{H}_{\rm s}^{k+1} - \vec{H}_{\rm s}^k - \vec{H}_{\rm p}^k )&=0~, \label{Hboundary_sm_n}
\end{align}
while for the shell-shell interfaces at $r=r_i$ with $2 \leq i<k$ 
\begin{align}
\vec{r} \times (\vec{E}_{\rm s}^{i+1} + \vec{E}_{\rm p}^{i+1} - \vec{E}_{\rm s}^{i} - \vec{E}_{\rm p}^{i})&=0~, \label{Eboundary_ss}\\ 
\vec{r} \times (\vec{H}_{\rm s}^{i+1} + \vec{H}_{\rm p}^{i+1} - \vec{H}_{\rm s}^{i} - \vec{H}_{\rm p}^{i})&=0~.\label{Hboundary_ss}
\end{align}
The boundary conditions for the core-shell interface at $r=r_1$ remain Eqs.~\eqref{Eboundary_sc} and \eqref{Hboundary_sc}.

Before we identify the hybridization scenario from the expansion coefficients emerging
from these boundary conditions we recall that the Mie approach works with the 
expansion coefficients whereas the hybridization scenario is an energetic picture 
focusing on energies (frequencies) for which the denominators of the expansion coefficients 
vanish. In deriving the energetic picture of the hybridization scenario from the expansion 
coefficients the question arises therefore on which coefficient one should focus. It turns 
out that any coefficient becoming resonant in the considered parameter regime can be taken. 
A $k$-layered sphere containing $k$ subsystems is described by the expansion
coefficients $c_n^i, b_n^{i+1}, d_n^i$, and $a_n^{i+1}$ with $i=1,\dots,k$. To extract 
the hybridization scenario we focus on $d_n^{i}$. It becomes resonant for small objects and 
is thus appropriate for our purpose.     

Our guiding principle is again the mathematical structure of the diagonal elements
of a matrix resolvent describing hybridized energy levels. Instead of two energy levels, 
appropriate for the core-shell particle, we now have to consider a resolvent for 
$k$ energy levels. Generalizing the approach described in Sect.~\ref{Theory} we thus 
have to rewrite $d_n^{i}$ in a form resembling the diagonal elements of the matrix resolvent
\begin{align}
 G(E) =
 \begin{pmatrix}
 \varepsilon_1 -E & V_{21}&0 & \dots &0 \\
 V_{21}^*&\varepsilon_2-E & V_{32}&\ddots & \vdots  \\
 0 & V_{32}^* & \ddots &\ddots& 0\\
 \vdots & \ddots &\ddots& \ddots&  V_{kk-1}\\
 0&\dots&0& V_{kk-1}^* &\varepsilon_k-E
 \end{pmatrix}^{-1}
 \; 
\label{Gmatrix}
\end{align}
describing a chain of energy levels $\varepsilon_i$ with nearest neighbor interactions $V_{ii-1}$,
where $i=1,\cdots, k$ and $V_{10}=V_{k+1,k}=0$. Using renormalized perturbation theory
they are given by~\cite{EC71} ($|i\rangle$ denotes the basis states)
\begin{widetext}
\begin{align}
	G_{i}(E)=\braket{i|G(E)|i}=\cfrac{1}{g^{-1}_{i}(E) -
			\cfrac{|V_{i+1i}|^2}{ g^{-1}_{i+1}(E) - 
				\cfrac{|V_{i+2i+1}|^2}{  
					\ddots - \cfrac{|V_{kk-1}|^2}{g_k^{-1}(E)  } }  } -
		\cfrac{|V_{ii-1}|^2}{ g^{-1}_{i-1}(E) - 
			\cfrac{|V_{i-1i-2}|^2}{   
				\ddots - \cfrac{|V_{21}|^2}{g_1^{-1}(E)  } }  }}~,
	\label{GFi}
\end{align}
\end{widetext}
where $g_i(E)=(E-\varepsilon_i)^{-1}$ is the resolvent of the $i$th non-interacting 
energy level. Its pole gives thus the energy of this level. The poles of $G_i(E)$ on the 
other hand give the $k$ excitation energies of the interacting system. Once the coefficient
$d_n^{i}$ is in this form its interpretation along the lines of the hybridization scenario 
is obvious as in the case of a core-shell particle.

Choosing all components nonmagnetic, that is, setting $\mu_i=1$ for $i \leq k+1$,
the coefficient $d_n^{i}$ resulting from the boundary conditions can be indeed rewritten as
\begin{widetext}
	\begin{align}
		d_n^{i}&=\cfrac{X_n^{k}(x_k)X_n^{k-1}(x_{k-1}) \cdots X_n^i(x_i) / 
[ \tilde{E}_n^k(x_1,\cdots,x_k)\tilde{E}_n^{k-1,k}(x_1,\cdots,x_k)\cdots\tilde{E}_n^{i+1,i+2,\dots,k}(x_1,\cdots,x_k)]  }{E_n^{i}(x_i) - 
			\cfrac{ X_n^{{\rm cp} i}(x_1,\dots,x_{i}) Z_n^{i+1}(x_{i+1})}{E_n^{i+1}(x_{i+1}) - 
				\cfrac{ X_n^{ {\rm cp} i+1}(x_1,\dots,x_{i+1}) Z_n^{i+2}(x_{i+2})}{   
					\ddots -\cfrac{  X_n^{{\rm cp} k-1}(x_1, \dots, x_{k-1}) Z_n^{k}(x_k)}{E_n^{k}(x_k)}  } }-
				\cfrac{ X_n^{{\rm cp} i-1}(x_1,\dots,x_{i-1}) Z_n^{i}(x_i)}{E_n^{i-1}(x_{i-1}) - 
					\cfrac{ X_n^{ {\rm cp} i-2}(x_1,\dots,x_{i-2}) Z_n^{i-1}(x_{i-1})}{   
						\ddots -\cfrac{  X_n^{{\rm cp} 1}(x_1) Z_n^{2}(x_2)}{E_n^{1}(x_1)}  } } }  
      \label{dnpi} \, , 
	\end{align}
\end{widetext}
that is, in a form closely resembling Eq.~\eqref{GFi}. Obviously $E_n^{i}$ 
and $X_n^{ {\rm cp}i-1}Z_n^{i}$ play, respectively, the role of $g_i^{-1}$ and $|V_{ii-1}|^2$. 
The functions $X_n^{i}$ and $\tilde{E}_n^{i,i+1,\dots,k}$ in the numerator, having no direct analogue,
are irrelevant for our purpose as they are not involved in the identification of resonance 
frequencies and coupling strengths. For better readability we again present the definitions of 
all these functions when they enter the discussion of the physical content
of Eq.~\eqref{dnpi}. 

The analogy between a chain of $k$ hybridized energy levels and the optical response of a $k$-layered 
sphere can be perhaps best seen by expressing Eq.~\eqref{dnpi} as a diagonal element of a tridiagonal 
matrix. Suppressing the size parameters Eq.~\eqref{dnpi} can be written as
\begin{align}
d_n^i=& \frac{X_n^{k} X_n^{k-1} \cdots X_n^i}
{\tilde{E}_n^{k}\tilde{E}_n^{k-1,k}\cdots \tilde{E}_n^{i+1,i+2,\dots,k}}
\braket{i|M|i}~,
\end{align}
with 
\begin{widetext}
\begin{align}
M=&\begin{pmatrix}
E_n^1 & \sqrt{X_n^{{\rm cp}1}Z_n^2}&0 & \dots &0 \\
\sqrt{X_n^{{\rm cp}1}Z_n^2}&E_n^2 &\sqrt{X_n^{{\rm cp}2}Z_n^3}&\ddots & \vdots  \\
0 & \sqrt{X_n^{{\rm cp}2}Z_n^3} & \ddots &\ddots& 0\\
\vdots & \ddots &\ddots& \ddots& \sqrt{X_n^{{\rm cp}k-1}Z_n^k}\\
0&\dots&0& \sqrt{X_n^{{\rm cp}k-1}Z_n^k} &E_n^k
\end{pmatrix}^{-1}
~.
\label{Mmatrix}
\end{align}
\end{widetext}

Comparing~\eqref{Mmatrix} with~\eqref{Gmatrix} it becomes clear that the optical response 
of a $k$-layered particle can be interpreted in terms of $k$ hybridized subsystems with 
individual resonances determined by ($i=1,\cdots,k$)
\begin{align}
	E^i_n(x_i)&=N_i \xi'_n(N_{i+1} x_i) \psi_n(N_{i} x_i)  \nonumber\\
	& - N_{i+1} \xi(N_{i+1} x_i) \psi'_n(N_{i} x_i) =0 \label{Eni}
\end{align}
and hybridization strengths given by $X_n^{ {\rm cp}i-1}(x_1,\dots,x_{i-1})Z_n^{i}(x_i)$.
Notice in Eq.~\eqref{Eni} the implicit definition of the function $E_n^{i}(x_i)$ which 
also appears in Eq.~\eqref{dnpi}. Clearly, Eq.~\eqref{Eni} is the resonance condition of an 
homogeneous particle with size parameter $x_i$ and refractive index $N_i$ embedded in a 
medium specified by $N_{i+1}$.

We now turn to the hybridization strength of the subsystem $i$, which in contrast to the 
core-shell particle interacts with two adjacent subsystems. The interaction with the $(i-1)$th and 
the $(i+1)$th subsystem is given by 
$X_n^{ {\rm cp}i-1}(x_1, \dots, x_{i-1})Z_n^{i}(x_i)$ and $X_n^{ {\rm cp} i}(x_1, \dots, x_{i})Z_n^{i+1}(x_{i+1})$, 
respectively. Here,
\begin{align}
	Z^i_n (x_i) &=N_{i} \xi'_n(N_{i+1} x_i) \xi_n(N_i x_i)  \nonumber \\
	& -N_{i+1} \xi_n(N_{i+1} x_i) \xi'_n(N_{i} x_i) \label{Zni}
\end{align}
for $i=2, \dots, k$ and
\begin{align}
	X_n^{ {\rm cp}i}(x_1, \dots ,x_i)&=X_n^{ {\rm p} i}(x_i) \nonumber \\
	 &+ \frac{ X_n^{ {\rm cp}i-1}(x_1,\dots,x_{i-1})U_n^{i}(x_i)}{E_n^{1,\cdots, i-1}(x_1,\dots, x_{i-1})}\; , \label{Xcpi} 
\end{align}
for $i=1, \dots, k-1$ with $X_n^{ {\rm cp}0}=0$. The functions in the definition of $X_n^{ {\rm cp}i}$  are 
\begin{align}
	X^{ {\rm  p} i}_n(x_i) &=N_i \psi'_n(N_{i+1} x_i) \psi_n(N_{i} x_i) \label{Xpi} \nonumber \\
	&- N_{i+1} \psi_n(N_{i+1} x_i) \psi'_n(N_{i} x_i) \; , \\
	U^i_n(x_i) &= N_{i+1}  \psi_n(N_{i+1} x_i) \xi'_n(N_{i} x_i)  \nonumber \\
	&- N_i  \psi'_n(N_{i+1} x_i) \xi_n(N_{i} x_i) \; ,
\end{align}
and
\begin{align}
	E_n^{1,\cdots, i}&(x_1,\dots,x_i)=  \nonumber \\
	 & E_n^{i}(x_i) + \frac{ X_n^{ {\rm cp}i-1}(x_1,\dots,x_{i-1}) Z_n^{i}(x_i)}{E_n^{1,\cdots, i-1}(x_1,\dots,x_{i-1})}~,
\label{RR1}
\end{align}
where $E_n^{i}(x_i)$ is defined in Eq.~\eqref{Eni}. The function $X_n^{\rm cpi}$ depends on the 
size parameters $x_1, \cdots , x_{i}$. Its superscript ${\rm cpi}$, standing for composed, propagating and 
$i$th shell, distinguishes it from the function $X_n^{\rm pi}$ which depends only on the size parameter 
$x_i$. The recursion~\eqref{RR1} generates the second continued fraction in the denominator 
of Eq.~\eqref{dnpi}.

Of interest is also under what conditions the $i$th subsystem of the stratified sphere can be isolated 
and how this affects the expansion coefficient $d_n^i$.
The decoupling is a two step process. In a first step, the interaction with the outer subsystems is turned 
off by choosing the outer shells and the embedding medium to be identical to the $(i+1)$th shell, that is, 
setting $N_{j}=N_{i+1}$ for $ i+1 < j \leq k+1$, which leads to $Z_n^{j}(x_{j})=0$ for $i+1 \leq j \leq k$.
In a second step, we turn off the interaction between the inner subsystems by choosing the refractive indices 
of the internal layers $j$ for $j<i$ and the refractive index of the $i$th layer to be the same. Hence, 
$N_{j}=N_{i}$ for $j<i$, so that $X_n^{ {\rm cp} i-1}(x_1,\dots, x_{i-1})=0$. To analyze how the isolation
of the $i$th subsystem affects the expansion coefficient $d_n^{i}$ we need also to look at the functions
\begin{align}
X_n^i(x_i) &=N_{i} \psi_n(N_{i+1}x_i) \xi^\prime_n(N_{i+1}x_i) \nonumber \\
&- N_i \psi^\prime_n(N_{i+1}x_i) \xi_n(N_{i+1} x_i) \label{Xni}
\end{align}
and 
\begin{align}
\tilde{E}_n^{i,i+1,\dots,k}&(x_1,\dots,x_k)=\nonumber \\
& E_n^{i}(x_i)  + \frac{X_n^{ {\rm cp}i}(x_1,\dots,x_i) Z_n^{i+1}(x_{i+1})}{\tilde{E}_n^{i+1,i+2,\dots,k}(x_1,\dots,x_k)}\; , \label{Eij}
\end{align}
entering the numerator in Eq.~\eqref{dnpi}.
The functions $E_n^i(x_i)$, $Z_n^i(x_i)$ and $X_n^{ {\rm cp}i}(x_1,\dots,x_{i})$ are 
defined in Eqs.~\eqref{Eni}, \eqref{Zni}, and \eqref{Xcpi}, respectively. Note, the 
recursion~\eqref{Eij} defines the first continued fraction in the denominator of \eqref{dnpi}. 
Due to the decoupling, that is, the particular identification of refractive indices, the 
interaction term $X_n^{ {\rm cp}i}(x_1,\dots,x_i) Z_n^{i+1}(x_{i+1})$ 
in Eq.~\eqref{Eij} vanishes leading to $\tilde{E}_n^{i,\cdots,k}(x_1,\cdots,x_k)=E_n^i(x_i)$.
In addition $X_n^j(x_j)/E_n^j(x_j)=1$ for $j>i$. Thus, Eq.~\eqref{dnpi} becomes 
\begin{align}
	d_n^{i} &= \frac{X_n^{i}(x_{i})}{E_n^{i}(x_{i})}\;  \label{dp_homo} 
\end{align}
with $X_n^{i}(x_{i})$ and $E_n^{i}(x_{i})$ given by Eqs.~\eqref{Xni} and~\eqref{Eni}.

Due to the decoupling $d_n^{i}$ should be attached to the penetrating fields inside an 
homogeneous particle characterized by $N_{i}$ embedded in a medium with the refractive index 
$N_{i+1}$. Indeed, looking at the penetrating coefficient of an homogeneous sphere given by 
Eq.~\eqref{dn_homo} and substituting $N_m\to N_{i+1}$, $N_2\to N_{i}$, and $x_2 \to x_i$ makes~\eqref{dn_homo} 
identical to \eqref{dp_homo}. Hence, if $N_j=N_{i}$ for all $j<i$ and $N_j=N_{i+1}$ for all $j>i+1$ 
the $k$-layered particle is reduced to an homogeneous particle with radius $r_i$ embedded in a medium, 
whereby the used materials are described by $N_{i}$ and $N_{i+1}$, respectively.

As in the case of the core-shell particle, the original expansion coefficients of a 
$k-$layered sphere~\cite{B85,SQ94,X05,PP09} reduce in 
the respective limits also to the coefficients of the subsystems. Due to the rewriting we demonstrate 
however that the subsystems are always virtually present. The rewriting identifies thus the limiting 
subsystems as basic building blocks whose coupling yields the unique optical response of 
the composite object.



\end{document}